\documentclass[a4paper]{jpconf}
\usepackage{graphicx}
\begin{document}

\title{Superfluid-Mott glass quantum multicritical point on a percolating lattice}

\author{Martin Puschmann$^{1,2}$ and Thomas Vojta$^1$  }

\address{$^1$ Department of Physics, Missouri University of Science \& Technology, Rolla, MO 65409, USA }
\address{$^2$ Institute of Physics, Technische Universit\"at Chemnitz, 09107 Chemnitz, Germany}
\ead{vojtat@mst.edu}

\begin{abstract}
We employ large-scale Monte Carlo simulations to study a particle-hole symmetric site-diluted
quantum rotor model in two dimensions. The ground state phase diagram of this system features two distinct quantum phase transitions
between the superfluid and the insulating (Mott glass) phases. They are separated by a multicritical point. The
generic transition for dilutions below the lattice percolation threshold is driven by quantum fluctuations
while the transition across the percolation threshold is due to the geometric fluctuations of the lattice.
We determine the location of the multicritical point between these two transitions and find its critical behavior.
The multicritical exponents read $z=1.72(2)$, $\beta/\nu=0.41(2)$, and $\gamma/\nu=2.90(5)$ .
We compare our results to other quantum phase transitions in disordered systems, and we discuss experiments.
\end{abstract}

\section{Introduction}
\label{sec:Intro}

Randomly diluted quantum many-particle systems feature two kinds of fluctuations at zero temperature.
First, geometric fluctuations due to bond and/or site dilution
of the lattice lead to a percolation problem with a geometric phase transition at the corresponding percolation
threshold \cite{StaufferAharony_book91}. Second, quantum fluctuations are caused by non-commuting operators
in the many-particle Hamiltonian. Combining both often leads to rich ground-state phase diagrams with
several different quantum phase transitions.

Prototypical examples of such behavior can be found in magnetic systems (for a review see,
e.g., Ref.\ \cite{VojtaHoyos08b}).
The diluted transverse-field Ising model displays long-range order for dilutions $p$ up to the percolation
threshold $p_c$ provided that the transverse-field $h_x$ is below the critical field $h_c(p)$. For
dilutions above the percolation threshold, long-range order is not possible since the system consists
of decoupled finite-size clusters. As a result, the $p$-$h_x$ ground state phase diagram displays
two quantum phase transitions, separated by a multicritical point at $(p_c,h^\ast)$
\cite{Harris74b,Stinchcombe81,Santos82,SenthilSachdev96,HoyosVojta06}.
The generic transition across $h_c(p)$ for $0<p<p_c$ is driven by quantum fluctuations while the
percolation transition across $p_c$ for $h_x<h^\ast$ is driven by the geometric criticality of the lattice.
Other quantum magnets behave similarly. The diluted square lattice Heisenberg quantum antiferromagnet
is long-range ordered at the percolation threshold \cite{Sandvik01,Sandvik02b}. If the quantum fluctuations
are strengthened, for example via an interlayer coupling in a bilayer quantum antiferromagnet, a phase diagram
with a multicritical point appears, as in the transverse-field Ising case
\cite{Sandvik02,VajkGreven02,VojtaSchmalian05b,VojtaSknepnek06}.

In the present paper, we investigate the interplay of geometric and quantum fluctuations in a system
of interacting bosons on a diluted lattice that undergoes a quantum phase transition between a superfluid
and an insulating ground state. Experimental applications include
Josephson junction arrays \cite{ZFEM92,ZEGM96}, granular superconducting films
\cite{HavilandLiuGoldman89,HebardPaalanen90}, helium absorbed in vycor \cite{CHSTR83,Reppy84},
as well as doped quantum magnets in high fields
\cite{OosawaTanaka02,HZMR10,Yuetal12}. In the presence of disorder, the superfluid phase and the Mott
insulator are separated by a ``glassy'' quantum Griffiths phase \cite{Griffiths69,ThillHuse95,Vojta06}
in which rare regions of superfluid order exist in an insulating bulk system. For generic
disorder (without special symmetries), this intermediate phase is the Bose glass, a compressible gapless
insulator \cite{FisherFisher88,FWGF89,PPST09}.
If the Hamiltonian is particle-hole symmetric even in the presence of disorder, the intermediate phase
between the superfluid and the Mott insulator is instead the so-called Mott glass (or random-rod glass)
\cite{GiamarchiLeDoussalOrignac01,WeichmanMukhopadhyay08} which is an incompressible gapless insulator.

While the quantum phase transition between superfluid and Bose glass has been studied in great detail,
the quantum phase transition between superfluid and Mott glass has attracted comparatively less
attention. Early numerical work in two dimensions using quantum Monte Carlo simulations \cite{ProkofevSvistunov04,SLRT14}
and a strong-disorder renormalization group \cite{IyerPekkerRefael12} produced inconclusive and partially
contradicting results for the critical behavior of this transition. To resolve this issue, we recently
performed large-scale Monte Carlo simulations of a site-diluted two-dimensional quantum
rotor model with particle-hole symmetry \cite{Vojtaetal16}.
Similar to the quantum magnets discussed above, this system features two distinct quantum phase transitions.
For dilutions $p$ below the percolation threshold $p_c$ of the lattice, we find a
superfluid-Mott glass transition characterized by a universal (dilution-independent)
critical behavior. The transition across the lattice percolation threshold $p_c$
falls into a different universality class.

In the present paper, we briefly summarize these results and then focus on the multicritical point
separating the generic quantum phase transition ($p<p_c$) from the percolation quantum phase transition
across the lattice percolation threshold. The paper is organized as follows.
We introduce in Sec.\ \ref{sec:Model} the quantum rotor Hamiltonian and its mapping onto a classical XY model with
columnar disorder. Section \ref{sec:MC} is devoted to the Monte Carlo method and the data analysis techniques.
Results for the generic transition, the percolation transition, and the multicritical point are
discussed in Sec.\ \ref{sec:Results}. We conclude in Sec.\ \ref{sec:Conclusions}.

\section{Diluted quantum rotor model}
\label{sec:Model}

The quantum rotor Hamiltonian is a prototypical model of interacting bosons. We consider a diluted square-lattice array
of superfluid grains, coupled by Josephson junctions. The Hamiltonian reads
\begin{equation}
 H = \frac U 2 \sum_i \epsilon_i (\hat n_i - \bar n_i)^2 -J\sum_{\langle ij \rangle} \epsilon_i \epsilon_j \cos(\hat \phi_{i}-\hat \phi_j)
\label{eq:Hamiltonian}
\end{equation}
where $\hat \phi_i$ is the phase operator of site (grain) $i$, and $\hat n_i$ is the number operator canonically conjugate
to the phase.
The sum over $\langle ij \rangle$ comprises pairs of nearest neighbors on the square lattice.
The parameters $U$ and $J$ are the charging energy and the Josephson coupling, respectively.
$\bar n_i$ is the offset charge at site $i$. Site dilution is introduced via the quenched random variables
$\epsilon_i$ which are independent of each other. They can take the values 0 (vacancy) with probability $p$ and 1
(occupied site) with probability $1-p$.

For generic values of the offset charges $\bar n_i$ or generic non-integer filling, the Hamiltonian (\ref{eq:Hamiltonian})
is not particle-hole symmetric, and the intermediate phase between superfluid and Mott insulator is a Bose glass.
To preserve the particle-hole symmetry, we set all offset charges $\bar n_i=0$ and fix the particle number $\langle \hat n \rangle$
at a large integer value. In this case, the disorder introduced by the site dilution is off-diagonal, and the intermediate
phase is a Mott glass.

The qualitative behavior of the particle-hole symmetric site-diluted quantum rotor model can be easily understood
\cite{FWGF89,WeichmanMukhopadhyay08,Vojtaetal16}.
If the Josephson energy $J$ dominates over the charging energy $U$ and if the site dilution $p$
is below the lattice percolation threshold $p_c$, the ground state is a long-range ordered superfluid.
Long-range order can be destroyed either by increasing the charging energy (which enhances the quantum fluctuations)
or by raising the dilution beyond the percolation threshold (where the lattice decomposes into
disconnected finite-size clusters).

Because we plan to perform Monte Carlo simulations using a highly efficient classical cluster algorithm,
we now map the quantum rotor Hamiltonian (\ref{eq:Hamiltonian}) onto a classical model. In the presence of
particle hole symmetry, the mapping leads to a classical XY model on a cubic lattice
\cite{WSGY94}. The classical Hamiltonian reads
\begin{equation}
 H_{\rm cl} = -J_s\sum_{\langle i,j\rangle, t}
\epsilon_i\epsilon_j\mathbf{S}_{i,t}\cdot\mathbf{S}_{j,t}
-J_\tau \sum_{i,t}\epsilon_i\mathbf{S}_{i,t}\cdot\mathbf{S}_{i,t+1}~.
\label{eq:Hcl}
\end{equation}
Here $i$ denotes a position in two-dimensional real space and $t$ is the ``imaginary time'' coordinate.
The dynamical variable $\mathbf{S}_{i,t}$ at each lattice site $(i,t)$ is an O(2) unit vector.
Note that the vacancy positions do not depend on the imaginary time coordinate $t$. The classical XY
model (\ref{eq:Hcl}) therefore features columnar defects, i.e., the disorder is perfectly correlated in
the imaginary time direction.
The values $J_s$ and $J_\tau$ of the interaction energies as well as the ``classical''
temperature $T$ are determined by the mapping from the original quantum rotor
Hamiltonian (the classical temperature differs from the real physical temperature which is zero).
As we are interested in the critical behavior which is expected to be universal, the precise
values of $J_s$ and $J_\tau$ are not important. Consequently, we fix $J_s=J_\tau=1$
and tune the classical XY model (\ref{eq:Hcl}) through its phase transitions by varying either
the classical temperature $T$ or the dilution $p$.

In the absence of dilution, the Hamiltonian (\ref{eq:Hcl}) represents the usual three-dimensional XY model. The clean
two-dimensional superfluid-Mott insulator quantum critical point is therefore in the
three-dimensional classical XY universality class which features a correlation length exponent
$\nu \approx 0.6717$ \cite{CHPV06}. The stability of the clean critical behavior against disorder
is governed by the Harris criterion  \cite{Harris74} $d\nu >2$. Here, $d$ is the number of dimensions
in which there is randomness; for columnar defects in a cubic lattice this implies $d=2$.
The Harris criterion is therefore violated, and the three-dimensional clean XY critical point is
unstable against columnar defects.

\section{Monte Carlo method}
\label{sec:MC}

To determine the critical behavior of the classical XY Hamiltonian (\ref{eq:Hcl}), we perform
Monte Carlo simulations using the highly efficient Wolff cluster algorithm \cite{Wolff89} which greatly
reduces the critical slowing down close to the phase transition. In addition to the Wolff updates,
we also perform conventional Metropolis updates \cite{MRRT53} to equilibrate small disconnected
clusters of lattice site that can occur at higher dilutions $p$ and may be missed by the Wolff
algorithm.

Our simulations cover several dilutions $p=0$, 1/8, 1/5, 2/7, 1/3, 9/25 and the lattice percolation
threshold $p_c=0.407253$.
Linear sizes range from $L=10$ to 150 in the two space directions
and $L_\tau=6$ to 1792 in the imaginary time direction. All observables are averaged over 10,000
to 50,000 disorder configurations, depending on system size. Each sample is equilibrated for
100 full Monte Carlo sweeps, followed by a measurement period of 500 sweeps with a measurement
taken after every sweep. (A full Monte Carlo sweep consists of a Metropolis sweep and a Wolff sweep,
the latter is defined as a number of
cluster flips such that the total number of flipped spins equals the number of
lattice sites.) The choice of rather short runs for a large number of disorder configurations
helps reducing the overall statistical error of the data, as discussed in Refs.\
\cite{VojtaSknepnek06,BFMM98,BFMM98b,SknepnekVojtaVojta04,ZWNHV15}. To eliminate systematic biases
in observables
stemming from the short simulation runs, we use improved estimators (see, e.g., the appendix of Ref.\ \cite{ZWNHV15}).

We analyze the Monte Carlo data by finite-size scaling. As the columnar disorder in the classical XY model
breaks the symmetry between the space and imaginary time directions, observables generally do not scale in the same way
with the system sizes $L$ and $L_\tau$. We consider, for example, the average Binder cumulant
\begin{equation}
\label{eq:Binder_av} g_{\rm av}=\left[ 1-\frac{\langle |\mathbf{m}|^4\rangle}{3\langle
|\mathbf{m}|^2\rangle^2}\right]_{\rm dis}~.
\end{equation}
Here,  $\left[\ldots\right]_{\rm dis}$ refers to the disorder average, $\langle\ldots\rangle$
denotes the Monte Carlo average for each sample, and
$\mathbf{m} = (1/N)\sum_{i,\tau}\mathbf{S}_{i,\tau}$ is the order parameter (with $N$ being the
number of lattice sites).
The finite-size scaling form of the Binder cumulant,
\begin{equation}
g_{\rm av}(r,L,L_\tau) = X_{g_{\rm av}}(rL^{1/\nu},L_\tau/L^z)~,
\label{eq:Binder_FSS}
\end{equation}
depends on two independent arguments. Here, $r=(T-T_c)/T_c$ is the  distance
from criticality, $z$ is the dynamical critical exponent,
and $X_{g_{\rm av}}$ is the dimensionless scaling function.\footnote{We have assumed
conventional power-law dynamical scaling rather than activated scaling for which the
scaling combination would be $\ln L_\tau / L^\psi$ (with $\psi$ the tunneling exponent), in agreement with the general
classification of disordered critical points put forward in Refs.\ \cite{Vojta06,VojtaSchmalian05}.}

In the absence of a value for the dynamical exponent $z$, the usual analysis
of the Binder cumulant (which identifies the critical point with the crossing
of the curves for different $L$) breaks down because the correct sample shapes are not known.
We therefore follow the method devised in Refs.\ \cite{VojtaSknepnek06,SknepnekVojtaVojta04,GuoBhattHuse94,RiegerYoung94}.
The Binder cumulant has a maximum as function of $L_\tau$ at fixed $L$ and $T$.
At the critical temperature, the peak position $L_\tau^{\rm max}$ scales as $L^z$.
This allows us to use an iterative approach that finds the value of $z$
together with the critical point.
The method begins with a guess of the dynamical exponent and the corresponding
sample shapes. The approximate crossing of the  $g_{\rm av}$ vs.\ $T$ curves for these shapes results in
an estimate for the critical temperature $T_c$. We then consider $g_{\rm av}$ as a function of $L_\tau$
for fixed $L$ at this temperature. The peak positions $L_\tau^{\rm max}$ provide improved estimates for
the optimal sample shapes. We iterate these steps three or four times to converge the values of $T_c$ and
$z$ with reasonable accuracy.

Once the optimal sample shapes are known (which fixes the scaling combination $L_\tau/L^z$),
the finite-size scaling analysis of other thermodynamic
observables such as the order parameter $m$ and its susceptibility $\chi$ proceeds normally. Their
finite-size scaling forms read
\begin{eqnarray}
m &=& L^{-\beta/\nu} X_m(rL^{1/\nu},L_\tau/L^z)~, \label{eq:m_FSS}\\
\chi &=& L^{\gamma/\nu} X_\chi(rL^{1/\nu},L_\tau/L^z) \label{eq:chi_FSS}
\end{eqnarray}
where $X_m$ and $X_\chi$
are dimensionless scaling functions, and $\beta$ and $\gamma$ are the order parameter
and susceptibility critical exponents, respectively. Correlation lengths in space
and imaginary time directions are calculated, as usual, from the second moment
of the spin correlation function \cite{CooperFreedmanPreston82,Kim93,CGGP01}.
The finite-size scaling forms of the  reduced correlation lengths are given by
\begin{eqnarray}
\xi_s/L &=& X_{\xi_s}(rL^{1/\nu},L_\tau/L^z)~, \label{eq:xis_FSS}\\
\xi_\tau/L_\tau &=&  X_{\xi_\tau}(rL^{1/\nu},L_\tau/L^z)~. \label{eq:xit_FSS}
\end{eqnarray}

\section{Results}
\label{sec:Results}

\subsection{Phase diagram}
\begin{figure}
\centerline{\includegraphics[width=10cm]{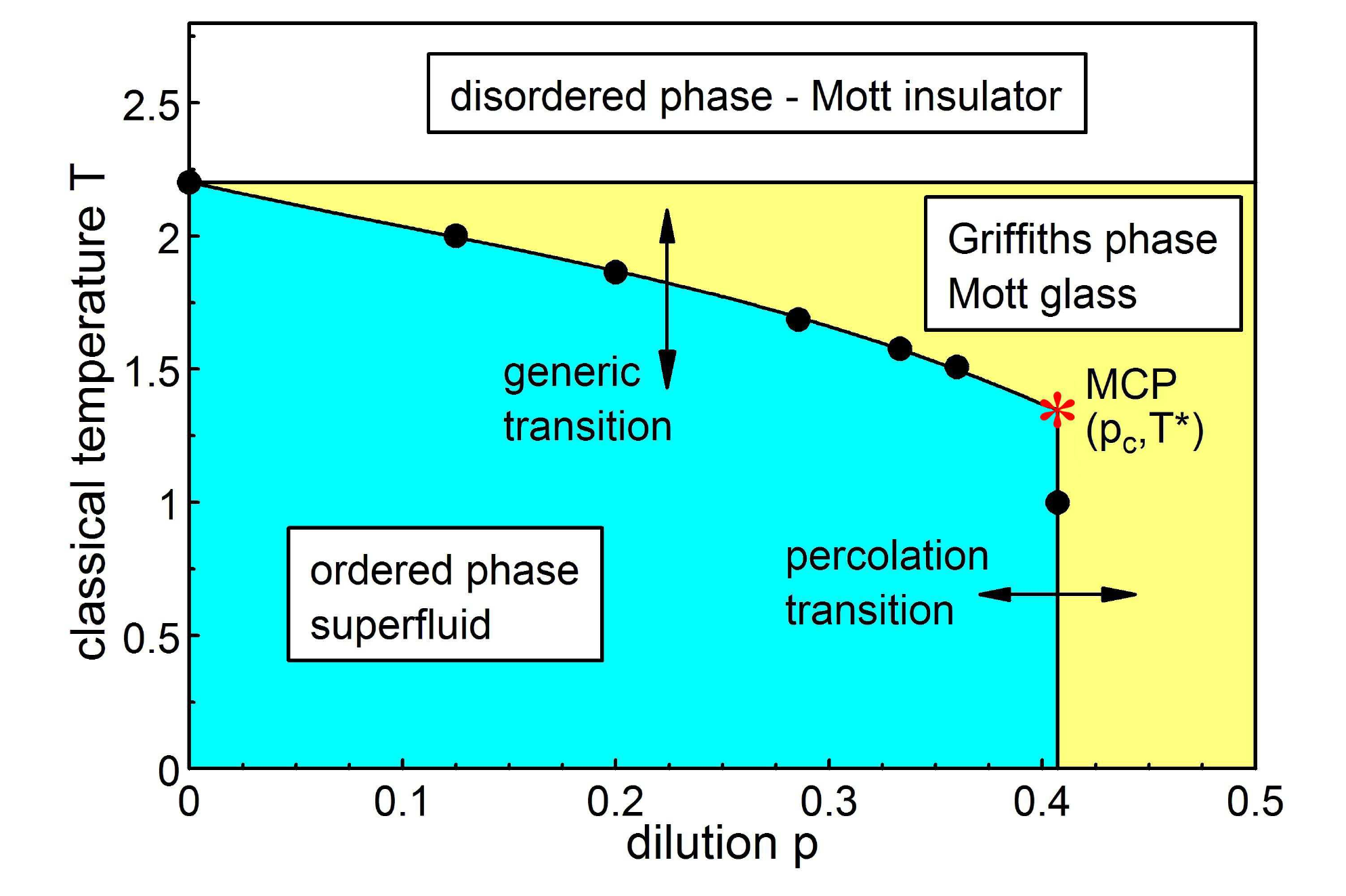}}
\caption{(Color online) Phase diagram of the $(2+1)$-dimensional XY model (\ref{eq:Hcl})
as a function of dilution $p$ and classical temperature $T$.
The black dots mark the numerically determined critical points on the generic and
percolation phase boundaries. The red star denotes the multicritical point (MCP)
at $p=p_c=0.407253$ and $T=T^\ast=1.337$
that separates the generic and percolation transitions. The lines
are guides for the eye only.}
\label{fig:pd}
\end{figure}

The phase diagram of the classical Hamiltonian (\ref{eq:Hcl}) in the dilution-(classical) temperature plane
that arises from our Monte Carlo simulations is presented in Fig.\ \ref{fig:pd}.
The generic phase boundary (for $p<p_c$) was determined in Ref.\ \cite{Vojtaetal16} by
performing simulations at dilutions $p=0$, 1/8, 1/5, 2/7, 1/3, and 9/25. In that work, we
also confirmed the location of the percolation phase transition by carrying out simulations
at $p=p_c$ and $T=1.0$.  The position of the multicritical point follows from our present
simulations (see below); it is located at $p=p_c=0.407253$ and $T=T^\ast=1.337$.

Qualitatively, this phase diagram is very similar to the phase diagrams of the diluted quantum
magnets discussed in the introductory section. Long-range (superfluid) order is restricted to
dilutions $p\le p_c$ and sufficiently weak quantum fluctuations (which are represented by the classical
temperature $T$ after the mapping of the rotor Hamiltonian onto the classical XY model).
Importantly, as for the quantum magnets, long-range order survives on the critical percolation cluster
(at $p=p_c$) up to a nonzero classical temperature $T^\ast$. This causes the existence of two distinct
phase transitions (generic and percolation) separated by the multicritical point that is the
focus of the present paper. Note that the disordered (insulating) phase can be decomposed into
two regions. If the classical temperature is below the transition temperature $T_c(0)$ of the undiluted system,
large spatial regions without vacancies can show local superfluid order even if the bulk is insulating.
This is the Griffiths phase of the transition, i.e., the Mott glass. For classical temperatures above
$T_c(0)$, local order is impossible; the system is thus a conventional Mott insulator.

\subsection{Generic and percolation transitions}
In this subsection, we briefly summarize the critical behaviors of the generic and percolation transitions.
To find the critical exponents of the generic transition, we analyzed \cite{Vojtaetal16} Monte Carlo data
for dilutions $p=1/8$, 1/5, 2/7, 1/3, and 9/25. By including correction-to-scaling terms that account for deviations
from pure power-law behavior, we could fit all these data using universal, dilution-independent exponents.
The values of the dynamical exponent $z$, the scale dimension $\beta/\nu$ of the order parameter, the corresponding
susceptibility exponent $\gamma/\nu$, and the correlation length exponent $\nu$
are listed in Table \ref{table:exponents}.
\begin{table}
\caption{Critical behavior of the superfluid-Mott glass quantum phase transitions. The exponents of the
generic ($p<p_c$) transition were found numerically in Ref.\ \cite{Vojtaetal16}. The exponents of the percolation
transition were determined from a scaling theory \cite{VojtaSchmalian05b} and numerically confirmed in Ref.\ \cite{Vojtaetal16}.
Their values are exact. The multicritical exponents are the result of the present paper.}
\label{table:exponents}
\begin{center}
\renewcommand*{\arraystretch}{1.2}
\begin{tabular*}{10cm}{l @{\extracolsep{\fill}} lll}
\br
Exponent            & generic      & percolation    & multicritical     \\
\mr
$z$                 & 1.52(3)      &  91/48        &       1.72(2)  \\
$\beta/\nu$         & 0.48(2)      &  5/48         &       0.41(2)\\
$\gamma/\nu$        & 2.52(4)      &  59/16        &       2.90(5)    \\
$\nu$               & 1.16(5)      &  4/3          &           \\
\br
\end{tabular*}
\end{center}
\end{table}
Other exponents can be found using various scaling relations.

The critical behavior of the transition across the lattice percolation threshold was analyzed earlier
by means of a scaling theory \cite{VojtaSchmalian05b} that relates its exponents to the classical
percolation exponents (which are known exactly in two space dimensions). The resulting exponent
values are listed in Table \ref{table:exponents} as well. Our Monte Carlo data for
the percolation transition (taken at $p=p_c$ and $T=1.0$) agree with these predictions \cite{Vojtaetal16}.

\subsection{Multicritical point}
We now turn to the main topic of this paper, viz., the multicritical point that separates the
generic and percolation transitions. To find the multicritical temperature $T^\ast$ and the dynamical
exponent $z$, we employ the iterative procedure described in Sec.\ \ref{sec:MC}. It consists of two
kinds of Monte Carlo runs: (i) runs right at $T_c$ for systems of different $L_\tau$ for each $L$.
The position $L_\tau^{\rm max}$ of the maximum of the Binder cumulant $g_{\rm av}$ yields the
optimal sample shapes and the dynamical exponent. (ii) In the second type of runs, we vary the
temperature but only use the optimal sample shapes. The crossing of the $g_{\rm av}$ vs.\ $T$
curves gives an (improved) estimate of $T_c$.

\begin{figure}[b]
\centerline{\includegraphics[width=10cm]{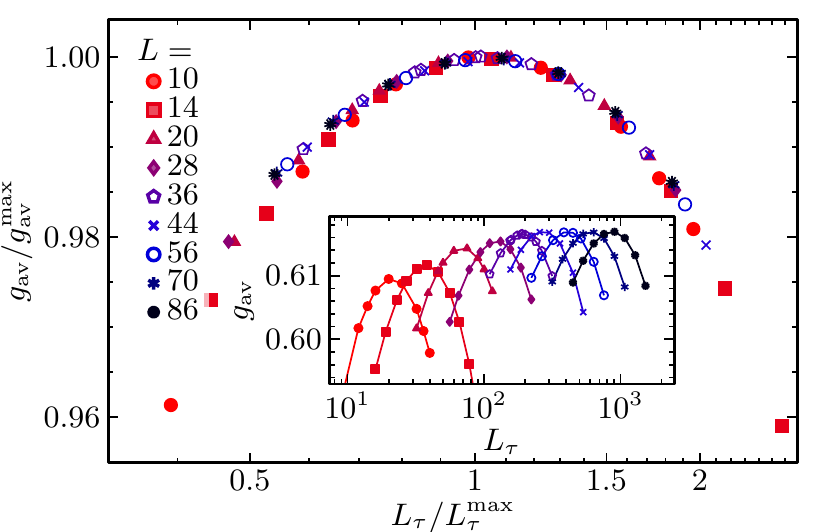}}
\caption{(Color online) Binder cumulant $g_{\rm av}$ as function of $L_\tau$ for several $L$
        at the multicritical temperature $T^\ast=1.337$. The relative statistical error of $g_{\rm av}$ is between 0.05\% and 0.08\%.
        Inset: Raw data $g_{\rm av}$ vs.\ $L_\tau$.
        Main panel: Scaling plot $g_{\rm av}/g_{\rm av}^{\rm max}$
        vs.\ $L_\tau/L_\tau^{\rm max}$.}
\label{fig:g_L_Lt}
\end{figure}

Results of the first type of runs are presented in Fig.\ \ref{fig:g_L_Lt}. 
It shows the Binder
cumulant $g_{\rm av}$ as function of $L_\tau$ for several $L=10$ to 86 after the iteration has
converged to a multicritical temperature of $T^\ast=1.337$.
(The error $\Delta T^\ast \approx 0.003$ to 0.005 is estimated heuristically from the quality
of the crossings of the $g_{\rm av}$ vs.\ $T$ curves for optimally shaped samples.)
The figure demonstrates that the maximum Binder cumulant $g_{\rm av}^{\rm max}$ for each of the
curves does not depend on $L$ (at least for the larger $L$), it takes a universal value of
about 0.6171(3). This is the behavior expected
at a (multi)critical point. The deviations for smaller $L$ can be attributed
to finite-size effects, i.e., to corrections to scaling.
The scaling plot presented in the main panel
of Fig.\ \ref{fig:g_L_Lt} confirms that $g_{\rm av}$ fulfills the scaling form (\ref{eq:Binder_FSS})
with high precision (slight deviations for small $L$ and $L_\tau$ stem, again, from corrections to scaling).

\begin{figure}
\centerline{\includegraphics[width=10cm]{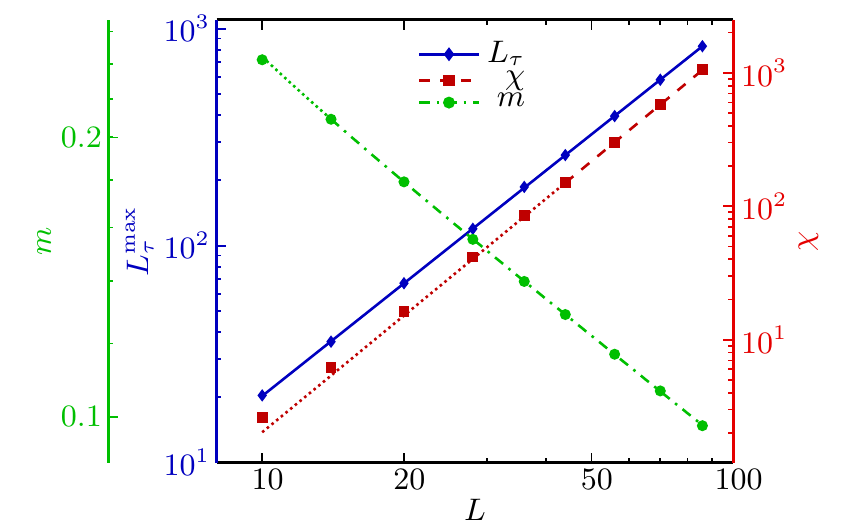}}
\caption{(Color online) Double-logarithmic plot of $L_\tau^{\rm max}$, $m$, and $\chi$ vs.\ $L$ for samples of the optimal shape
        at the multicritical temperature $T^\ast=1.337$. The statistical errors are well below the symbol size.
        The lines represent power-law fits (see text). They are dotted in the regions not included in the fit.
        }
\label{fig:scaling}
\end{figure}

Once the optimal shapes are known, the dynamical critical exponent $z$ can be found by analyzing
the dependence of $L_\tau^{\rm max}$ on $L$, as shown in Fig.\ \ref{fig:scaling}.
The data can be fitted well using a pure power-law dependence $L_\tau^{\rm max} \sim L^z$ giving a value
$z=1.725(3)$. The statistical errors of $L_\tau^{\rm max}$ are determined from 1000 synthetic data sets. Each set
is obtained from the original data by adding Gaussian random noise corresponding the uncertainties of the
original data. The power-law fit is of good quality, it yields $\tilde \chi^2\approx 0.74$.
[The reduced sum of squared errors of the fit (per degree of freedom)
is denoted by $\tilde \chi^2$ to distinguish it from the order parameter susceptibility $\chi$.]
In addition to the statistical error, $z$ also has an error stemming from the uncertainty in $T^\ast$.
To estimate it, we repeat the $L_\tau^{\rm max}$ vs.\ $L$ analysis at the higher temperature
$T^\ast+ 0.01$, significantly further away from $T^\ast$ than our estimated $T^\ast$ error.
This leads to a shift of $z$ of about 0.025. Our final estimate for the dynamical critical exponent
thus reads $z=1.72(2)$.

The critical exponents $\beta/\nu$ and $\gamma/\nu$ can be determined from the $L$-dependence of the
order parameter and the susceptibility, respectively, for samples of optimal shape at the multicritical
temperature. The corresponding plots are also shown in Fig.\ \ref{fig:scaling}. The order parameter data
for $L>20$ can be fitted well by a pure power law $m \sim L^{-\beta/\nu}$ giving $\beta/\nu=0.411(1)$
with $\tilde \chi^2\approx 0.99$. If the leading correction-to-scaling term is included via
$m = aL^{-\beta/\nu}(1+bL^{-\omega})$, the fit range can be extended to all $L>10$, yielding an almost
unchanged exponent value $\beta/\nu=0.406(4)$ with $\tilde \chi^2\approx 1.26$. The largest source of error
is again the uncertainty in $T^\ast$. To estimate it, we repeat the analysis at $T^\ast\pm \Delta T^\ast$
with $\Delta T^\ast=0.003$.
This leads to shifts of $\beta/\nu$ of about 0.02. Our final estimate therefore is $\beta/\nu=0.41(2)$.

The susceptibility data show stronger deviations from pure power-law behavior. A fit to
$\chi \sim L^{\gamma/\nu}$ works for sizes $L>36$ and yields $\gamma/\nu=2.905(4)$ with $\tilde \chi^2\approx 0.66$.
Extending the fit range to sizes $L>28$ gives $\gamma/\nu=2.880(3)$ but leads to unacceptably large  $\tilde \chi^2$ values. The fit becomes
unstable when a correction-to-scaling term is included via $\chi = aL^{\gamma/\nu}(1+bL^{-\omega})$. Due to these
instabilities $\gamma/\nu$ has a larger error.
Our final value is  $\gamma/\nu =2.90(5)$ where the error bar is estimated heuristically from the robustness
of the fit; it includes the error due to the uncertainty in $T^\ast$ obtained from repeating the analysis at $T^\ast\pm 0.003$.

Note that the multicritical exponents $\beta/\nu$, $\gamma/\nu$, and $z$  must fulfill the hyperscaling relation
$2\beta/\nu+\gamma/\nu=d+z$ where $d=2$ is the number of real space dimensions.
The Monte Carlo estimates $\beta/\nu=0.41(2)$, $\gamma/\nu=2.90(5)$, and
$z=1.72(2)$ fulfill this relation within their error bars. This gives
us confidence that they represent true asymptotic rather than effective critical
exponents.

\section{Conclusions}
\label{sec:Conclusions}

In conclusion, we have reported the results of large-scale Monte Carlo simulations of a
site-diluted particle-hole symmetric quantum rotor model in two dimensions. They were aimed at investigating
the quantum phase transitions between the superfluid and insulating (Mott glass) ground states
of disordered interacting bosons.
The ground state phase diagram in the dilution-quantum fluctuation plane features
two distinct quantum phase transitions, (i) the generic transition that occurs for
dilutions below the lattice percolation threshold and is driven by quantum fluctuations, and (ii)
the transition across the lattice percolation threshold which is driven by geometric
fluctuations. The two phase transition lines meet at a multicritical point at which
both quantum and geometric fluctuations become critical.

In the absence of disorder, the superfluid-Mott insulator transition belongs to the
three-dimensional classical XY universality class. Its correlation length exponent
$\nu\approx 0.6717$ violates the Harris criterion $d\nu > 2$. The critical exponents
of the diluted system are therefore expected to differ from the clean ones. This expectation is
fulfilled by our Monte Carlo results, summarized in Table \ref{table:exponents},
for the generic transition, the percolation transition, and the multicritical point.
In all cases, the transition is of conventional ``finite-disorder'' type
(featuring power-law dynamical scaling) rather than an infinite randomness critical point
\cite{Fisher92,MMHF00,HoyosKotabageVojta07,VojtaKotabageHoyos09}
or a smeared transition \cite{Vojta03a,HoyosVojta08}.

We have considered here the two-component quantum rotor Hamiltonian that is a
model of disordered interacting bosons. The analogous three-component quantum rotor
Hamiltonian which arises as an effective low-energy theory of a bilayer Heisenberg
antiferromagnet was studied in Refs.\ \cite{VojtaSknepnek06,SknepnekVojtaVojta04}.
Both systems have very similar phase diagrams, but their exponent values differ.
For example, the dynamical exponent $z$ of the two-component model considered here
takes the values 1.52 and 1.72 for the generic transition and the multicritical point,
respectively. The corresponding values for the three-component case are 1.31 and 1.54.
Other exponents show similar differences. Note, however, that the exponents of the
percolation quantum phase transition do not depend on the number of rotor components, as they
are determined by the geometric criticality of the lattice only.

It is also interesting to compare the influence of the space dimensionality. The one-dimensional
equivalent of the rotor Hamiltonian (\ref{eq:Hamiltonian}) undergoes a Kosterlitz-Thouless
quantum phase transition \cite{KosterlitzThouless73} in the two-dimensional classical XY universality class in the
absence of disorder. This transition fulfills the Harris criterion (as $\nu=\infty$),
weak disorder therefore does not change the critical behavior. At larger disorder, the character of
the transition does change even though the details of its critical behavior are still
controversially discussed (see, e.g., Refs.\ \cite{AKPR04,AKPR10,HrahshehVojta12,PolletProkofevSvistunov14}
and references therein). In contrast, the clean rotor model in three space dimensions
violates the Harris criterion because its correlation length exponent takes the
mean-field value 1/2. We therefore expect a scenario similar to the two-dimensional case
considered in the present paper.

Experimentally, there are several ways to realize diluted systems of interacting bosons.
These include granular superconductors (whose superconductor-insulator quantum phase transition
is believed to be bosonic in nature, at least in some systems), ultracold atoms (where
the dilution could perhaps be realized via mixtures of heavy and light atoms), or
certain magnetic systems such as diluted anisotropic
spin-1 antiferromagnets \cite{RoscildeHaas07}. The last example is particularly
promising because the required particle-hole symmetry appears naturally as
 a consequence of the up-down symmetry of the spin Hamiltonian in the absence
 of an external magnetic field.

\ack

This work was supported in part by the NSF under Grant Nos.\ DMR-1205803
and DMR-1506152.
M.P. acknowledges support by an InProTUC scholarship of
the German Academic Exchange Service during the early parts of this work.
We thank Daniel Arovas, Jack Crewse, Yury Kiselev,
Gil Refael, and Nandini Trivedi for helpful discussions.

\section*{References}

\bibliographystyle{iopart-num}
\bibliography{../00Bibtex/rareregions}

\end{document}